\documentclass{emulateapj}
\usepackage{graphicx}
\usepackage[space]{grffile}
\usepackage{latexsym}
\usepackage{textcomp}
\usepackage{longtable}
\usepackage{multirow,booktabs}
\usepackage{amsfonts,amsmath,amssymb}
\usepackage{natbib}
\usepackage{url}
\usepackage{hyperref}
\hypersetup{colorlinks=false,pdfborder={0 0 0}}
\newif\iflatexml\latexmlfalse
\usepackage[utf8]{inputenc}
\usepackage[english]{babel}

\newcommand{\code}[1]{\texttt{#1}}

\shorttitle{GADFLY: A pandas-based Framework for Analyzing GADGET Simulation Data}
\shortauthors{Jacob Hummel}

\begin{document}


\title{GADFLY:  A pandas-based Framework for Analyzing GADGET Simulation Data}


\author{Jacob Hummel}
\affil{Department of Astronomy, The University of Texas at Austin, TX 78712, USA}
\email{jhummel@astro.as.utexas.edu}




\begin{abstract}
We present the first public release (v0.1) of the open-source \textsc{gadget} Dataframe Library: \code{gadfly}.
The aim of this package is to leverage the capabilities of the broader python scientific computing ecosystem by providing tools for analyzing simulation data from the astrophysical simulation codes \textsc{gadget} and \textsc{gizmo} using \code{pandas}, a thoroughly documented, open-source library providing high-performance, easy-to-use data structures that is quickly becoming the standard for data analysis in python. 
\code{Gadfly} is a framework for analyzing particle-based simulation data stored in the HDF5 format using \code{pandas DataFrame}s. 
The package enables efficient memory management, includes utilities for unit handling, coordinate transformations, and parallel batch processing, and provides highly optimized routines for visualizing smoothed-particle hydrodynamics (SPH) datasets.
\end{abstract}

\bibliographystyle{apj}

\section{Introduction}
\label{sec:intro}

In the past decade, astrophysical simulations have increased dramatically in both scale and sophistication, and the typical size of the datasets produced has grown accordingly.  
However, the software tools for analyzing such datasets have not kept pace, such that one of the primary barriers to exploratory investigation is simply manipulating the data.  
This problem is particularly acute for users of the popular smoothed particle hydrodynamics (SPH) code \textsc{gadget} \citep{SpringelYoshidaWhite2001,Springel2005}.  
Both \textsc{gadget} and \textsc{gizmo} \citep{Hopkins2015}, which uses the same data storage format, are widely used to investigate a range of astrophysical problems; unfortunately this also leads to fractionation of the data storage format as each research group modifies the output to suit its needs.
This state of affairs has historically forced significant duplication of effort, with individual research groups separately developing their own unique analysis scripts to perform similar operations.

Fortunately, the issue of data management and analysis is not endemic to astronomy, and the resulting overlap with the needs of the broader scientific community and the industrial community at large provides a large pool of scientific software developers to tackle these common problems.
In recent years, this broader community has settled on python as its programming language of choice due to its efficacy as a `glue' language and the rapid speed of development it allows.  
This has led to the development of a robust scientific software ecosystem with packages for numerical data analysis like \code{numpy} \citep{VanderWaltColbertVaroquaux2011}, \code{scipy} \citep{JonesOliphantPeterson2001}, \code{pandas} \citep{McKinney2010}, and \code{scikit-image}; \code{matplotlib} \citep{Hunter2007} and \code{seaborn} for plotting; \code{scikit-learn} for machine learning, and statistics and modeling packages like \code{scikits-statsmodels}, \code{pymc}, and \code{emcee} \citep{Foreman-Mackeyetal2013}.

Python is quickly becoming the language of choice for astronomers as well, with the Astropy project \citep{Robitailleetal2013} and its affiliated packages providing a coordinated set of tools implementing the core astronomy-specific functionality needed by researchers. 
Additionally, the development of flexible python packages like \code{yt} \citep{Turketal2011} and \code{pynbody} \citep{Pontzenetal2013}, capable of analyzing and visualizing astrophysical simulation data from several different simulation codes, have greatly improved the ability of computational researchers to perform useful, insight-generating analysis of their datasets.

Recently, the scientific python community has begun to converge on the \code{DataFrame} provided by the high-performance \code{pandas} data analysis library as a common data structure. 
As a result, once data is loaded into a \code{DataFrame}, it becomes much easier to take advantage of the powerful analysis tools provided by the broader scientific computing ecosystem.
With this in mind, we present a \code{pandas}-based framework for analyzing \textsc{gadget} simulation data, \textsc{gadfly}: the \textsc{GAdget} DataFrame LibrarY.

In this paper we present the first public release (v0.1) of \code{gadfly}, which is available at \code{http://github.com/hummel/gadfly}. 
The framework design and organizational structure are outlined in Section \ref{sec:framework}, followed by a description of the included SPH particle rendering  in Section \ref{sec:vis}.  Our plans for future development are outlined in Section \ref{sec:future}, and a summary is provided in Section \ref{sec:summary}.

\begin{figure}[h!]
\begin{center}
\includegraphics[width=0.98\columnwidth]{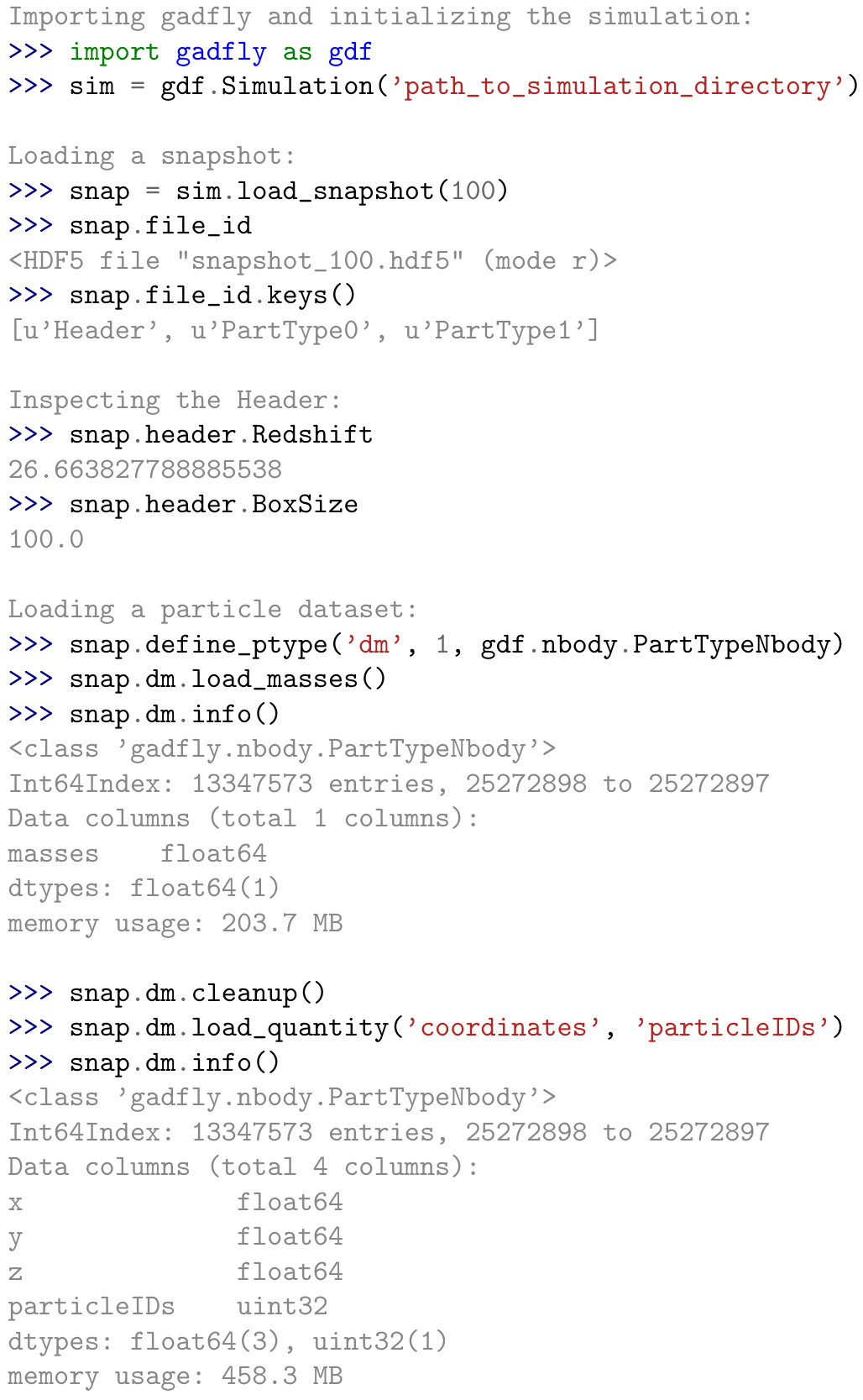}
\caption{\label{fig:usage_example}
Initializing a simulation, defining a \code{PartType} dataset, and loading data.%
}
\end{center}
\end{figure}

\section{A Framework built on pandas}
\label{sec:framework}

There are several motivations for building an analysis framework around the \code{pandas DataFrame}. 
The guiding principle underlying the design of this framework is to enable exploratory investigation.
This requires both intelligent memory management for handling out-of-core datasets, and a robust indexing system to ensure that particle properties do not become misaligned in memory.
Using  the \code{pandas DataFrame} as the primary data container rather than separate \code{numpy} arrays makes it much easier to keep different particle properties indexed correctly while still affording the flexibility to load and remove data from memory at will.
In addition, \code{pandas} itself is a thoroughly documented, open-source, BSD-licensed library providing high-performance, easy-to-use data structures and analysis tools, and has a strong community of developers working to improve it.  
More broadly, as \code{pandas} is becoming the de-facto standard for data analysis in python, doing so simplifies interoperability with the rest of the available tools.

\code{Gadfly} is designed for use with simulation data stored in the HDF5 format\footnote{\code{http://www.hdfgroup.org/HDF5}}.
While we otherwise aim to keep \code{gadfly} as general as possible, some assumptions about the storage format are necessary.
Each particle type is expected to be contained in a different HDF5 group, labeled \code{PartType0, PartType1}, etc; a \code{Header} group is also expected, containing metadata for the simulation snapshot as HDF5 attributes. 
All particles are expected to have the following fields defined: particle IDs, masses, coordinates, and velocities.  
SPH particles are additionally expected to have a smoothing length, density, and internal energy.  Additional fields not included in the original \textsc{gadget} specification, such as chemical abundances, are also supported.

Here, we provide an overview of the design and capabilities of the \code{gadfly} framework, including the \code{Simulation}, \code{Snapshot}, and \code{PartType DataFrame} objects at the core of \code{gadfly} (Section \ref{sec:hierarchy}), the usage of which is demonstrated in Figure \ref{fig:usage_example}.
Our approach to file access and intelligent memory management (Section \ref{sec:fileIO}), our handling of unit conversions (Section \ref{sec:units}) and coordinate transformations (Section \ref{sec:coords}), and the included utilities for parallel batch processing (Section \ref{sec:parallel}) are also described.

\begin{figure}[h!]
\begin{center}
\includegraphics[width=0.98\columnwidth]{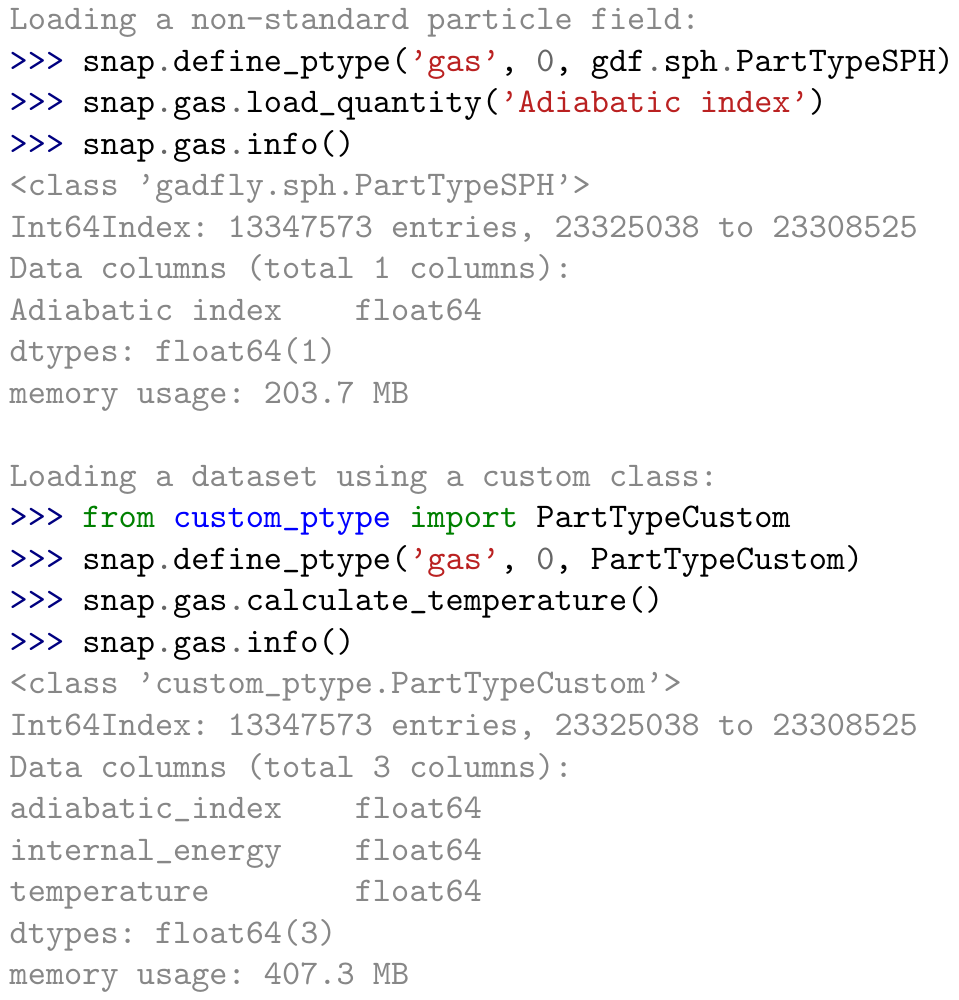}
\caption{\label{fig:custom_ptype}
Loading non-standard particle fields and defining custom PartType classes. Within \code{gadfly} this is straightforward.%
}
\end{center}
\end{figure}

\subsection{Organizational Structure}
\label{sec:hierarchy}

\subsubsection{\code{PartType} Dataframes}
\label{sec:df}
Data for each particle type (e.g., dark matter, gas, etc.) is stored in a separate \code{PartType} dataframe and indexed by particle ID number. 
Individual fields can be loaded into the dataframes and deleted at will, with coherent slicing across multiple data fields, courtesy of \code{pandas}.  
The base \code{PartType} objects, \code{PartTypeNbody} and \code{PartTypeSPH}, are standard \code{pandas} dataframes with additional functionality for loading standard \textsc{gadget} particle fields from HDF5, converting units, and performing coordinate transformations.  
As such, \code{PartType} dataframes retain all the capabilities of the \code{pandas.DataFrame} from which they inherit, and any operation that creates a new dataframe instance returns a standard \code{pandas} dataframe.  

Nonstandard particle fields (e.g., chemical abundances) can be easily loaded into \code{gadfly} as well; fields loaded in this manner simply lack automated unit conversion.
Alternatively, a custom dataframe class inheriting from \code{PartTypeNbody} or \code{PartTypeSPH} as appropriate can be defined to provide methods for loading both nonstandard particle fields and additional derived properties (e.g., temperature). An example of such a custom class---\code{PartTypeCustom.py}---is provided in the examples distributed with \code{gadfly}, and the usage of such a custom class is demonstrated in Figure \ref{fig:custom_ptype}.

\subsubsection{\code{Snapshot}}
\label{sec:snap}
Each \code{Snapshot} object represents a single HDF5 snapshot file on disk.  File access---provided by \code{h5py} \citep{h5py}---is handled via the \code{Snapshot} object, and the actual particle data, loaded as needed into the \code{PartType} dataframes described in Section \ref{sec:df}, is gathered here with each particle type contained in a separate \code{PartType} dataframe.  
The information contained in the \textsc{gadget} header is also maintained here in a \code{Header} object, along with access to the additional metadata and unit information stored in the relevant \code{Simulation} object.

\subsubsection{\code{Simulation}}
\label{sec:sim}
Metadata relevant to the simulation as a whole, such as filepaths and unit information, are stored in a \code{Simulation} object.  Initializing a \code{Simulation} object is the first step in any analysis using \code{gadfly} as this is where default values for all subsequent analysis are set, including locating all relevant snapshot files, choosing a coordinate system, and setting the field names of the default particle properties expected by \code{gadfly}.  \code{Snapshot}s are loaded via \verb|Simulation.load_snapshot()|, and the parallel batch processing functionality described in Section \ref{sec:parallel} is implemented at this level as well.

\subsection{Memory Management and File Access}
\label{sec:fileIO}
One of the primary goals of the \code{gadfly} project is to enable the analysis of large datasets on machines with limited memory.
Enabling this requires intelligent memory management, loading only the particle data of interest from the disk.
Fortunately the HDF5 protocol is well-suited to such non-contiguous file access, allowing not only individual data fields to be accessed independently, but also for  loading only select entries from the field in question.

\code{Gadfly} employs two complementary approaches to minimizing the memory footprint.
The first method requires definition of an optional refinement criterion, such as particles above a given density threshold.
The resulting `refined' index can then be used to select only the corresponding values from subsequent particle fields as they are loaded.
While this method efficiently minimizes I/O operations, it is fairly rigid, and attempting to load additional fields into a dataframe from which particles have been manually dropped will fail, as the particle indices will no longer match.  
As such, this approach is poorly suited to exploratory analysis where the proper refinement criterion may not be know {\it{a priori}} and is best suited for use in scripts where the analysis to be performed is well defined.

To mitigate the indexing issues that can arise in situations where data is loaded incrementally, \code{gadfly} performs an intermediate step, first loading new fields into a \code{pandas.Series} data structure, using the particle ID numbers as an index.  This allows \code{pandas} to properly align the particle fields, dropping any particles not in the existing \code{PartType} dataframe from the newly loaded field as it is appended.
This approach, which can be used in tandem with the refinement index, affords \code{gadfly} the flexibility needed to allow incremental manual refinement of the data stored in memory. 
Additional cuts can be made as subsequent fields are loaded, resulting in the selection of a precisely targeted primary dataset from which derived  properties (e.g., temperature) may be calculated, which serves to reduce computational overhead as well.

\begin{figure}[h!]
\begin{center}
\includegraphics[width=0.98\columnwidth]{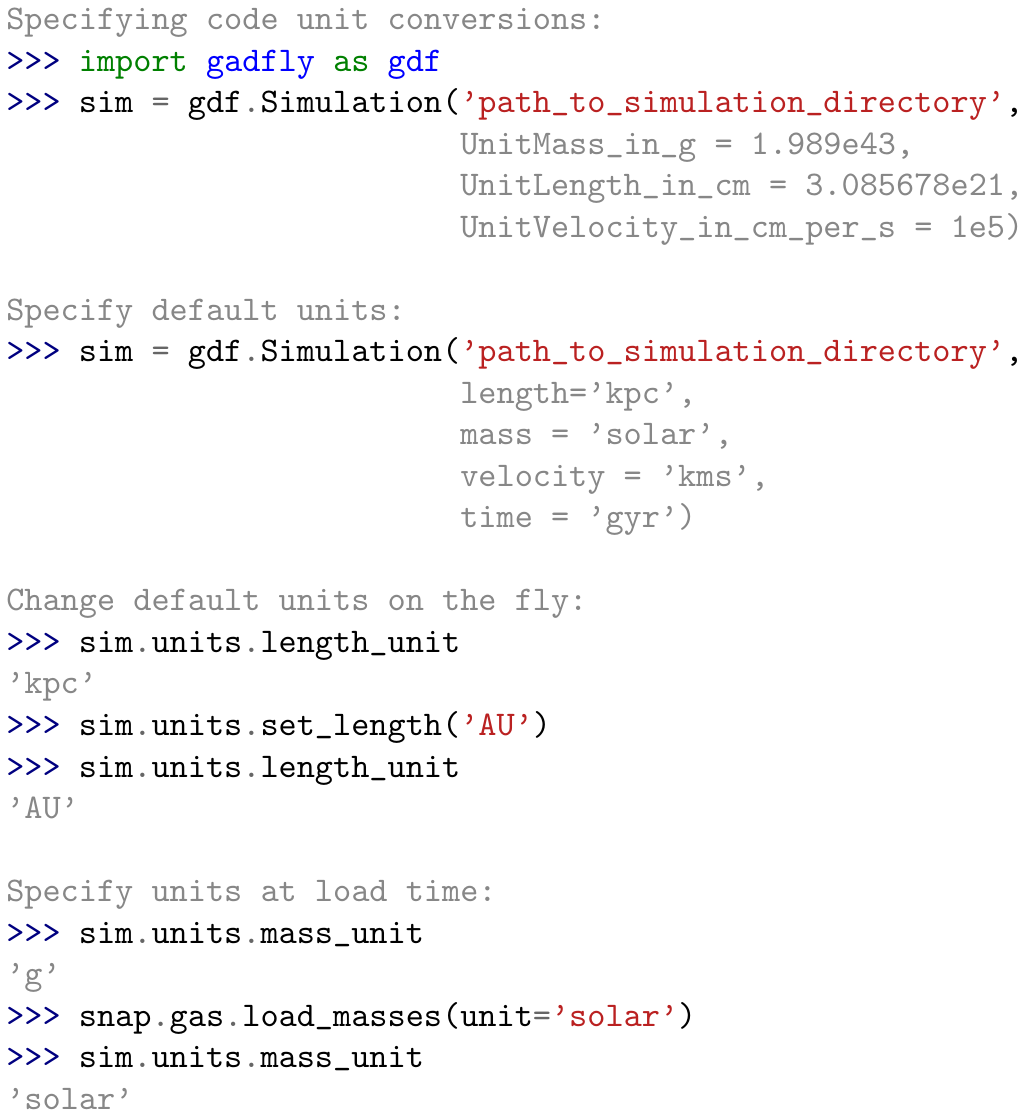}
\caption{\label{fig:units}
Unit handling in \code{gadfly}.%
}
\end{center}
\end{figure}

\subsection{Units}
\label{sec:units}
\code{Gadfly} implements a minimal unit-handling system for converting between the native code units in which \textsc{gadget} stores data and the physical units of interest to astronomers.  
However, \textsc{gadget}'s code units may be modified to suit the problem at hand, and therefore must be specified at initialization if they differ from the defaults listed in Table \ref{code_unit_defaults}.
\begin{table}[h!]
    \centering
    \caption{Default code units expected by \code{gadfly}.}
    \label{code_unit_defaults}
    \begin{tabular}{ll}
        \hline
        Unit & Conversion to \verb|cgs|\\
        \hline
        \verb|Mass| &  $1.989\times10^{43}\,$g\\ 
        \verb|Length| & $3.085678\times10^{21}\,$cm \\ 
        \verb|Velocity| & $1.0\times10^5\,$cm$\,$s$^{-1}$ \\ 
    \end{tabular} 
\end{table}
Conversion from code units to the physical units system of choice is handled at loading.
The default units for length, time, mass, etc. can be specified at initialization, along with whether to use physical or comoving coordinates (for cosmological simulations) and whether to factor the Hubble constant out of the reported units (i.e., units of Mpc vs. Mpc/$h$).
While defaults are set at startup, they can be modified at any time, either by calling \verb|units.set_length()|, which alters the default length unit for all subsequently loaded fields, or by calling a particle field's load function with the optional \code{units} keyword argument.  
Changing the units of an existing field can be done by simply reloading it; the field will remain properly indexed as described in Section \ref{sec:fileIO}. Examples of both approaches to unit conversions are shown in Figure \ref{fig:units}.

\subsection{Coordinate Transformations}
\label{sec:coords}
A full suite of coordinate transformation utilities is included in \code{gadfly}, with methods for converting between Cartesian, spherical, and cylindrical coordinates, performing linear coordinate translations, and rotations about arbitrary axes.  
These methods are both directly accessible as independent library functions, and via the \code{PartType} dataframe object using the \verb|.orient_box()| functionality.

\subsection{Parallel Batch Processing}
\label{sec:parallel}
Investigating the results of a simulation often requires running an identical analysis on many snapshots in order to study how the system changes with time.  
These operations are typically completely independent, and thus amenable to parallelization---so long as the machine being used has sufficient memory.
To aid in the parallelization of such analyses, \code{gadfly} includes utilities for performing parallel batch processing jobs by making use of python's \code{multiprocessing} module. 

One issue with doing operations such as this in parallel is that they are often IO-bound.
To mitigate the issues that arise when multiple processes attempt to read large amounts of data from disk simultaneously, \code{gadfly}'s parallelization utilities are designed to load the data needed for a given analysis serially in a separate thread from the main execution. 
As the necessary data from each snapshot file is loaded, execution is passed to a pool of worker processes which can then perform the remainder of the analysis in parallel.  An example of such a parallel batch processing script is included in the examples distributed with \code{gadfly}.

\begin{figure}[h!]
\begin{center}
\includegraphics[width=1\columnwidth]{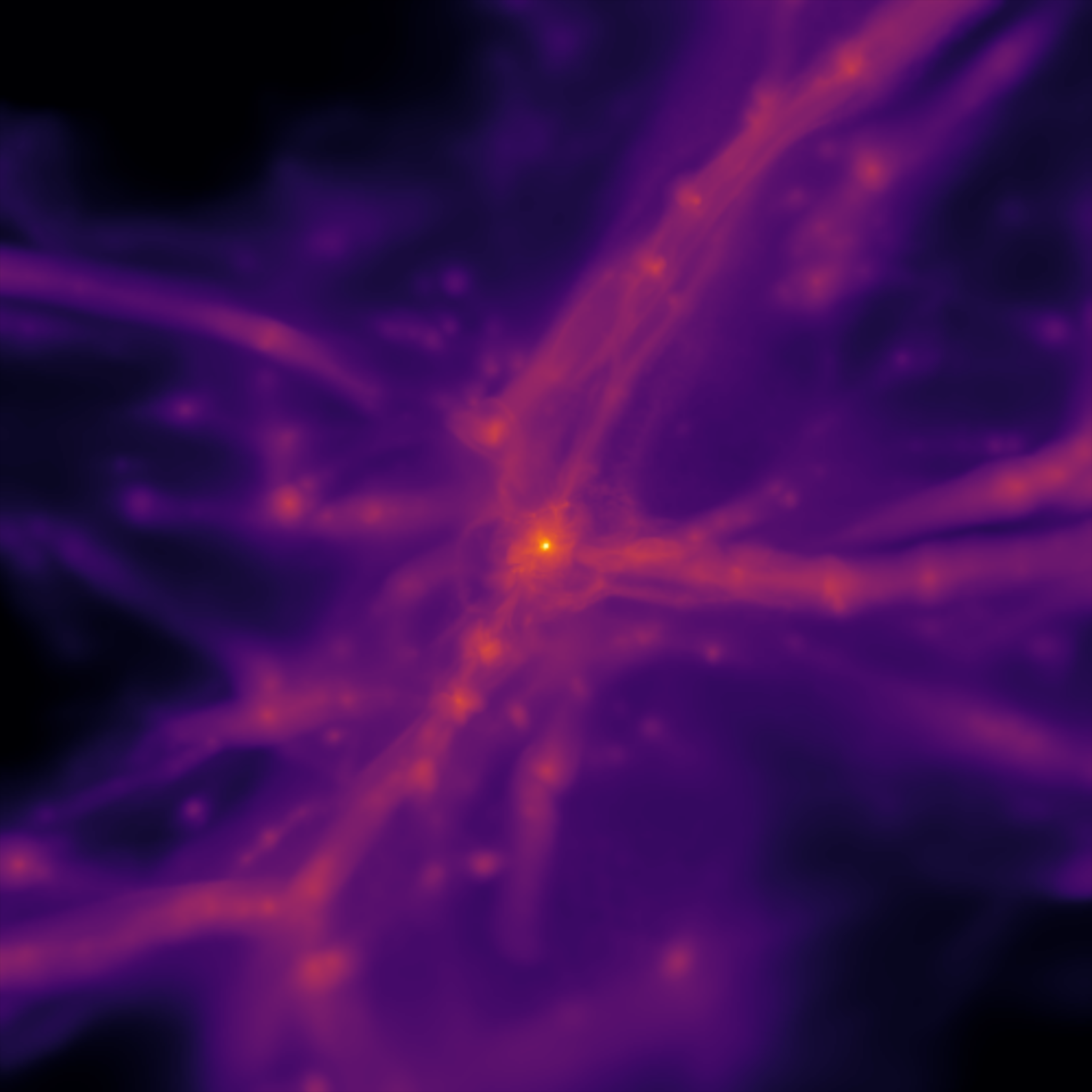}
\caption{\label{fig:vis}
Example of a visualization produced by \code{gadfly}.  This particular image shows the density distribution in a minihalo forming at $z=25$ on $1\,{\rm kpc}$ scales, and is from a simulation run for \citet{Hummeletal2015}.%
}
\end{center}
\end{figure}

\section{Visualization}
\label{sec:vis}
Visualization plays a major role in any analysis of simulation data, and the ability to directly visualize the spatial structure and evolution of a system is often crucial to generating insight.
While the guiding principle of the \code{gadfly} project is to avoid reinventing the wheel, instead relying on the tools developed by the broader scientific python community wherever possible, SPH particle rendering is one critical area where that broader ecosystem proves insufficient.  
To mitigate this shortcoming, \code{gadfly} includes a minimal library of visualization tools.

The particles in an SPH simulation are best thought of as fluid elements sampling the continuum properties of the gas they represent \citep{Lucy1977,GingoldMonaghan1977,Monaghan1992,Springel2010}.  They accomplish this by serving as Lagrangian tracers over which the continuum properties are interpolated using a smoothing kernel $W$. While it is possible to use alternative kernels, most modern SPH implementations (including \textsc{gadget}) utilize a cubic spline kernel \citep{Springel2014}: 
\begin{equation}
W(r,h_{\rm s}) =
     \begin{cases}
       1 - 6 \left( \frac{r}{h_{\rm s}} \right)^2 + 6 \left( \frac{r}{h_{\rm s}} \right)^3, & 0 \leq \frac{r}{h_{\rm s}} \leq \frac{1}{2}\\
       2 \left(1 - \frac{r}{h_{\rm s}}\right)^3, & \frac{1}{2} < \frac{r}{h_{\rm s}} \leq 1\\
       0, & \frac{r}{h} >  1,\\
     \end{cases}
\end{equation}
where $r$ is the radius and $h_{\rm s}$ is the characteristic width of the kernel, otherwise known as the smoothing length.  The physical density at any point, $\rho(\bf{r})$, is then represented by the sum over all particles
\begin{equation}
\rho({\bf r}) \simeq \sum_j m_j W({\bf r} - {\bf r}_j, h_{\rm s}),
\end{equation}
where $m_j$ is the mass of particle $j$, located at ${\bf r}_j$.
As such, creating visualizations that faithfully reproduce the actual density distribution requires performing this sum over all particles of interest; this can be quite computationally intensive depending on the number of particles involved and the desired resolution.
The SPH particle rendering algorithm included in \code{gadfly} performs this summation over two dimensions, producing a density-weighted projection. An example of such a visualization produced by \code{gadfly} is shown in Figure \ref{fig:vis}.

\code{Gadfly} includes three separate implementations of this algorithm, each of which is best suited to different conditions:
\begin{enumerate}
\item The primary implementation is derived from \citet{NavratilJohnsonBromm2007} and is written in \code{C}, parallelized using OpenMP, and wrapped in python using \code{scipy.weave}.  This method must be locally compiled, and will fail if the python interpreter cannot locate a \code{C} compiler, or if the requisite libraries are not installed.  However, it its the most powerful implementation, best suited to rendering many particles, and to machines with many processors available.
\item A second implementation makes use of \code{numba} \citep{LamPitrouSeibert2015} to perform just-in-time (JIT) compilation of pure python code using the LLVM compiler infrastructure \citep{LattnerAdve2004}.  The resulting serial routine is highly optimized, providing performance within a factor of two of the parallel method on a 16-core machine.  As such, this method is preferable on smaller machines with fewer processors, and for visualizations including fewer particles where the additional overhead associated with the parallel implementation is significant.
\item The final implementation is a pure python routine included only for situations where the other methods have unmet dependencies. This implementation is over 500 times slower than the others, and as such is suitable only to visualizing small numbers of particles, or as a last resort.
\end{enumerate}

These methods are all available as independent library functions, and can be used both with particle data in a \code{PartType} dataframe or independently from the rest of the \code{gadfly} framework, with data stored in \code{numpy} arrays.  
Future versions of \code{gadfly} will greatly simplify the visualization of data stored in a \code{PartType} dataframe through the use of a \code{.visualize()} method (see Section \ref{sec:future}).

\section{Future Development}
\label{sec:future}
The \code{gadfly} package is under active development, and in addition to incremental improvements of the existing functionality, a few significant upgrades are planned for future releases.  
First and foremost, the current \code{gadfly} units system is fairly inflexible, with limited support for additional units.  
For the next release we plan on replacing the existing units system, instead employing the far more versatile \code{astropy.units} framework \citep{Robitailleetal2013} as the backend for handling unit conversions. 
Future releases will also more seamlessly integrate \code{gadfly}'s visualization tools with the rest of the framework.  
At the moment the user is required to pass each required field to the visualization methods individually.
We intend to integrate this functionality directly into the \code{PartType} dataframe, allowing the user to simply call \code{PartType.visualize()} and be presented with a default density projection of the SPH particles currently in the dataframe, similar to the \code{.plot()} functionality of a standard \code{pandas.DataFrame}.

In the longer term, the \code{dask}\footnote{\code{http://dask.pydata.org}} project presents an intriguing option for handling very large out-of-core datasets through the use of blocked algorithms and task scheduling.
\code{Dask} manages this by mapping high-level \code{numpy}, \code{pandas}, and list operations on large datasets to many operations on smaller chunked datasets that can fit in memory.  
Future releases of \code{gadfly} may migrate the data structure on which the \code{PartType} dataframe is built from the \code{pandas.DataFrame} to \code{dask.dataframe} to take advantage of this functionality.

\section{Summary}
\label{sec:summary}
In this paper we have presented the first public release of \code{gadfly}.  We have described the framework's structure, which is designed around three core abstractions: the \code{PartType} dataframe (Section \ref{sec:df}), the \code{Snapshot} object (Section \ref{sec:snap}), and the \code{Simulation} object (Section \ref{sec:sim}).  Additional functionality includes intelligent memory management and file access (Section \ref{sec:fileIO}), basic unit handling (Section \ref{sec:units}), coordinate transformations (Section \ref{sec:coords}), utilities for parallel batch processing (Section \ref{sec:parallel}), and SPH particle visualization (Section \ref{sec:vis}).

\code{Gadfly} is fully open-source, is released under the MIT License, and can be downloaded and installed from its repository at \code{http://github.com/hummel/gadfly}.  Users can submit bug reports via GitHub, and if they know how to fix them, are welcome to submit pull requests.


\bibliography{bibliography/converted_to_latex.bib%
}

\end{document}